# Intrinsic Nanotwin Effect on Thermal Boundary Conductance in Bulk and Single-Nanowire Twinning Superlattices

Aaron Porter, Chan Tran, Frederic Sansoz[*]

*Mechanical Engineering Program, School of Engineering, University of Vermont, Burlington, Vermont 05405, USA*

[*]Corresponding author: frederic.sansoz@uvm.edu; (802) 656-3837

# ABSTRACT

Coherent twin boundaries form periodic lamellar twinning in a wide variety of semiconductor nanowires, and are often viewed as near-perfect interfaces with reduced phonon and electron scattering behaviors. Such unique characteristics are of practical interest for high-performance thermoelectrics and optoelectronics; however, insufficient understanding of twin-size effects on thermal boundary resistance poses significant limitations for potential applications. Here, using atomistic simulations and *ab-initio* calculations, we report direct computational observations showing a crossover from diffuse interface scattering to superlattice-like behavior for thermal transport across nanoscale twin boundaries present in prototypical bulk and nanowire Si examples. Intrinsic interface scattering is identified for twin periods $\geq$ 22.6 nm, but also vanishes below this size to be replaced by ultrahigh Kapitza thermal conductances. Detailed analysis of vibrational modes shows that modeling twin boundaries as atomically-thin 6*H*-Si layers, rather than phonon scattering interfaces, provides an accurate description of effective cross-plane and in-plane thermal conductivities in twinning superlattices, as a function of the twin period thickness.

## I. INTRODUCTION

Quantum-well superlattice structures composed of a periodic arrangement of nanolayers from two semiconductors with significant band-gap and mass mismatch, usually combine high thermoelectric power with low thermal conductivity at the nanoscale; two properties of primary interest for thermoelectric applications [1,2]. Evidence for a minimum of thermal conductivity when the superlattice period decreases below a critical thickness is well established [3-8]. Considerable research effort has been devoted to understanding the fundamental regime transition from incoherent to coherent in order to control thermal conductivity in superlattices. Recent experiments and atomistic simulations have shown that the incoherent heat conduction shift to coherent phonon transport occurs for small periods of 2-5 nm in thickness in semiconductor and epitaxial oxide superlattices with atomically–smooth interfaces [9-11]. Superlattices with rough interfaces, however, have significantly reduced coherent phonon transport behavior [9,12-14] and electron mobility [15], proving that a mix of perfectly coherent interfaces is highly desirable to achieve more efficient thermoelectric superlattices.

This article examines how periodic twin boundaries (TBs), which are naturally abundant coherent interfaces, can fundamentally influence the thermal transport behavior of bulk crystals and nanowires (NWs) in Si. These so-called *coherent twinning superlattices* with lamellar twinning running either perpendicular or parallel to the growth direction have been observed in various types of semiconductor NWs [16], including Si and SiC materials [17-22]. Given their high atomic-scale order and symmetry, TBs are considered as near-perfect interfaces for their reduced phonon and electron scattering characteristics compared to general grain boundaries [23-29]. As such the electronic band energies can change sharply across a TB, since this type of interface is a flat, atomically-thin hexagonal layer between two diamond-cubic grains [18]. For Si, the bandgap energy is smaller in the hexagonal phase (Si IV) than in the cubic one (Si I) [30]. Therefore, each added TB creates a potential difference forming a quantum well, which entails that twinning superlattices could effectively increase the Seebeck coefficient without sacrificing the electrical conductivity, unlike other conventional superlattices.

Yet our current understanding of heat transport in twinning superlattices remains incomplete. It has been shown that the Kapitza thermal conductance across a single interface can be one order of magnitude larger for TBs than other high-energy grain boundaries [23,27,28]. This implies that TBs offer less resistance to heat conduction by minimizing diffuse phonon scattering, although it has been shown that TBs selectively scatter phonons at particular frequencies [23]. The atomistic simulation study by Xiong *et al.* [7] has recently reported no temperature gap across TBs in nanotwinned bulk Si, suggesting that TBs play no intrinsic role on heat conduction. However, these authors have found a minimum of thermal conductivity in Si twinning superlattice NWs, attributed to extrinsic geometric effects from the zigzag surface morphology of the NWs. Likewise, the addition of one lengthwise TB parallel to the axis in Si NWs was found to produce negligible changes in thermal conductivity (<5%) by atomistic simulations [31]. Also, Dong *et al.* [32] have recently found a weak dependence of thermal boundary resistance from TBs in nanotwinned diamond by MD. However, these atomistic predictions contradict recent experiments in InP and Ge twinned NWs finding up to 50% reduction in thermal conductivity attributed to nanotwin effects [33,34], proving that the role of twin size on heat conduction is not clearly understood. Here, we report atomistic simulations showing evidence for an intrinsic nanotwin effect on thermal conductivity and twin boundary conductance in bulk and small-scale (NW) twinning superlattices in pure Si.

## II. METHODOLOGY

Diamond-cubic Si crystals containing nanoscale TBs were modeled with the MD simulation software LAMMPS [35] and OVITO [36] using the Stillinger-Weber potential [37]. This potential has shown good agreement for the prediction of phonon dispersion curves in bulk and NWs compared to *ab-initio* data [38], and has been used elsewhere in previous atomistic simulation studies of thermal transport in twinned Si [7,23]. The crystal orientation in adjacent grains was rotated by 60º about the $[1\bar{1}1]$ axis to form a Σ3(111) crystalline interface between each pair of grains. The models consisted of a constant twin period thickness $2\lambda$ separately varied from 1.88 nm to 45.2 nm, as shown in Fig. 1(a) and Fig. S1. Correspondingly, the TB density $1/\lambda$ ranged from 0 to 1.06 interfaces per nm. Perfect single crystals with either diamond-cubic (3*C*) or hexagonal structures (2*H*, 4*H*, and 6*H*) [30] were also simulated; see Supplemental Materials for details on atomistic modeling of these structures. The common neighbor analysis was used to differentiate between the twin and grain regions.

All simulations were performed in the microcanonical ensemble (constant number of particles *N*, volume *V* and energy *E*) with *T* = 300 K and 700 K, which was realized by using a Berendsen thermostat [39]. The time step was 1 fs. The thermal conductivity was computed by non-equilibrium MD (NEMD) simulations [40] where "hot" and "cold" slabs were positioned at the center and on both sides of the box, respectively. The simulation box was divided into 100 slabs perpendicular to the axis. It was previously noted by Skye and Schelling [41] that NEMD simulations of thermal conductivity remains entirely classical despite the fact that a temperature of 300 K is below the Debye temperature of Si (650 K as predicted by the Stillinger-Weber potential). Here, a relaxation to zero stress for 20 ps was performed along the main axis of each model prior to applying a heat flux $J$. The latter was imposed by exchanging the velocity vectors between the hottest atom in the cold slab and the coldest atom in the hot slab in a way that produced an energy transfer between the hot and cold regions. A total of 1 million simulation steps were performed in order to obtain linear temperature profiles in the two intervening regions (heat flux was positive in one region and negative in the other). No significant change in temperature profiles was found for the last 500,000 steps, which indicated that a steady state was reached for all simulations. From the two computed profiles, an average temperature gradient $\partial T/\partial x$ (with $x$ the long axis coordinate) was used to obtain the effective thermal conductivity $\kappa = -J/(\partial T/\partial x)$. Here, simulations were repeated with different heat exchanges ($J = 9.5 - 18$ eV·ps$^{-1}$). Cross-plane and in-plane thermal conductivities, denoted in the following with superscript symbols $\perp$ and $\parallel$, respectively, were measured perpendicular and parallel to the interfaces along $[1\bar{1}1]$ and <112> crystal orientations, by changing the heat flux direction.

Scaling effects on thermal conductivity computed from MD simulations, particularly the differences between NEMD and equilibrium (EMD) methods, have received considerable attention in the literature [40,42,43]. For the NEMD method, a size-independent thermal conductivity is usually obtained by performing a linear extrapolation from multiple systems of different length [40]. It was argued that NEMD systematically leads to smaller bulk thermal conductivities than EMD due to non-linearity [42], whereas the latter has proved to show greater uncertainty in thermal conductivity predictions for inhomogeneous or superlattice systems with interface scattering [40,44]. In this study, however, one primary objective was to investigate the effects of TB spacing on thermal boundary conductance, for which NEMD is more adapted. By repeating the calculations with a different length $L$ varied from 45 nm to 180 nm, deviation from a linear scaling was found to be more pronounced in bulk Si models (Fig. S2) where periodic boundary conditions were imposed along the three spatial directions, than in Si NW models. Therefore, models of identical dimensions (9.8nm×9.8 nm×90.4 nm) were used to compare thermal conductivities in bulk materials, whereas different lengths ($L$=45nm, 90 nm and 150 nm) are presented for thermal conductivity in Si NWs.

Thermal boundary (Kapitza) conductance ($\sigma_K$) was calculated from the observed temperature discontinuity ($T_K$) at a TB interface between two grains, such as [45]

$$\sigma_K = \frac{J}{T_K}, \qquad (1)$$

with two high heat fluxes ($J$ = 24 eV·ps$^{-1}$ and 41 eV·ps$^{-1}$). For that purpose, the simulation box was divided into 288 slabs along the axis to compute the average temperature in each {111} close-packed planes. The temperature discontinuity for a single TB was calculated by taking the difference in temperature across the length $L_K$ on either side of the TB where $L_K$ was defined by the limit of the fitting lines shown in blue color in Fig. 1b. Mean value and standard deviation for the thermal boundary conductance were determined from measurements on 2 to 8 TBs depending on the system. Interfaces less than 5 nm from the hot and cold slabs were excluded from this evaluation to ensure the temperature discontinuities only belonged to the linear temperature profile.

Phonon density of states (PDOS) spectra from MD simulations were calculated by directly applying a Fourier transform on the atomic velocity distribution, which was found to be computationally more efficient than other methods involving either the velocity autocorrelation function [46] or direct dispersion curves calculations [47], while being mathematically equivalent. For the PDOS of a given subset of particles, the velocity components of each atom were stored every 20 fs for 163,840 timesteps, and a Fourier transform, normalized by the number of atoms present, was performed. The Nyquist frequency was 25 THz. A Wiener filter with 16 time windows [48] was used to reduce noise and average the PDOS curves. The frequency resolution after filtering was 0.1 THz.

First-principles calculations of PDOS for $3C$ and $2H$ Si were conducted from calculations of phonon dispersion curves obtained by density-functional-theory (DFT)-based *ab-initio* with the Quantum ESPRESSO code [49]. We used a norm-conserving pseudopotential for Si with the Perdew-Zunger (LDA) exchange-correlation non relativistic approach by van Barth and Car [50]. Periodic atomic structures used for these calculations are presented in Fig. S3 in Supplemental Materials.

## III. RESULTS AND DISCUSSION

### A. Twin boundary thermal conductance

A fully-periodic atomistic model of bulk nanotwinned Si oriented in the $[1\bar{1}\bar{1}]$ direction, shown in Fig. 1(a), was used to study the change of TB thermal conductance as a function of twin period thickness by imposing a constant heat flux through the interfaces at a temperature of 300 K. Fig. 1(b) shows a representative temperature profile for a bulk system with a twin period of 45.2 nm, the largest considered in this study, superimposed to the underlying atomic structure. Best fit lines used to calculate the temperature gradients for the grain region are shown in blue. We find the thermal conductivity of the grain region to be 22.1 W/mK, close to the value of 20.5 ± 1.0 W/mK for a twin-free 3*C* single crystal of same size that we computed with a smaller heat flux. The temperature profile displays a net temperature change across the interface due to interface phonon scattering. It is important to note that the temperature jump occurs over a finite length equal to 4 atomic layers ($L_K$ = 0.94 nm), suggesting that the heat conduction length across a single TB is small, but not zero. All thermal conductance presented in the following was evaluated over the same length $L_K$. First, we consider a bulk system of length $L$ = 90 nm with a heat flux of either 24 eV/ps or 41 eV/ps. Fig. 1(c) shows that the Kapitza boundary conductance for twin periods $2\lambda \geq 22.6$ nm is almost constant and independent on heat flux, $\sigma_K$ = 6.8 ± 1.6 GW/m²K. This thermal transport behavior is consistent with previous predictions for TB phonon scattering in crystalline Si. Aubry *et al.* [23] using a direct MD simulation at 500 K with the same interatomic potential obtained a boundary conductance of 10.2 GW/m²K in a Σ3(111) TB interface, and 0.80 GW/m²K in a more disordered Σ29(001) grain boundary, which is in good agreement with the present results. However, a striking feature in Fig. 1(c) is a dramatic climb in thermal boundary conductance up to 28 GW/m²K for twin periods $2\lambda <$ 22.6 nm, accompanied by a strong dependence on heat flux. Furthermore, increasing the system size to $L$ = 180 nm is found to have no effect on the Kapitza conductance in the interface scattering regime for twin period thickness $2\lambda \geq 22.6$ nm, but some effect for smaller twin periods, suggesting two distinct thermal transport behaviors in twinning superlattices as a function of twin size.

**B. Bulk cross-plane and in-plane thermal transport**

The cross-plane and in-plane thermal conductivities computed by using the entire temperature profile spanning over multiple interfaces in the superlattices are presented in Fig. 2 as a function of the TB density. Two different behaviors are predicted in Fig. 2(a) depending on the applied temperature. At 300 K, we find a 29% reduction in cross-plane thermal conductivity from 20.5 W/mK for a twin-free crystal to 14.6 W/mK for a superlattice with $2\lambda = 22.6$ nm. Subsequently, the cross-plane thermal conductivity for $2\lambda < 22.6$ nm (TB density $> 0.09$ nm$^{-1}$) increases linearly with the TB density, and thus the thermal conductivity at $2\lambda = 22.6$ nm represents an absolute minimum. By contrast, a previous atomistic simulation study [28] reported a continuous reduction of thermal conductivity up to TB densities $= 0.2$ nm$^{-1}$ in different nanotwinned FCC metals, which suggests that this minimum may only be observed in semiconductor twinning superlattices.

At 700 K, a 19% reduction in cross-plane thermal conductivity is found at a twin period thickness of 22.6 nm, which is smaller than at 300 K. Generally, an increase in temperature augments the Kapitza boundary conductance [51], which is in good agreement. For $2\lambda < 22.6$ nm, however, the thermal conductivity at 700 K continues to decrease linearly as a function of TB density, albeit more slowly than for $2\lambda \geq 22.6$ nm, which is opposite to the upward trend at 300 K. This type of temperature dependence predicted in twinning superlattices with very small twin periods is somewhat reminiscent of that observed experimentally in conventional superlattice materials with periods smaller than the phonon coherence length [9].

To confirm the hypothesis of a regime transition, the classical particle theory was used to calculate the effective cross-plane thermal conductivity $\kappa_{eff}^{\perp}$ by taking into account scattering interfaces at equal distance $\lambda$ [28,52]:

$$\kappa_{eff}^{\perp} = \left[\frac{1}{\kappa_{3C}^{\perp}} + \left(\frac{1}{\lambda}\right)(\sigma_K)^{-1}\right]^{-1}, \qquad (1)$$

where $\kappa_{3C}^{\perp}$ is the cross-plane thermal conductivity in the <111>-oriented twin-free 3C Si crystal. Fitting the simulation results in Fig. 2(a) with $\sigma_K$ = 7.0 GW/m²K and 8.0 GW/m²K at 300 K and 700 K, respectively, Eq. (1) shows an excellent match for $2\lambda \geq 22.6$ nm, but a significant departure from MD simulation results beyond this point.

Furthermore, the effective in-plane thermal conductivity $\kappa_{eff}^{\parallel}$ from MD simulations at 300 K represented in Fig. 2(b), shows no strict minimum, but the predicted value of 15.5 ± 0.4 W/mK is 20% smaller than 19.7 ± 0.3 W/mK in the twin-free crystal in the <112> direction, which only represents a 6% difference with the minimum cross-plane thermal conductivity in Fig. 2(a). This result is atypical given the known thermal conductivity anisotropy in conventional superlattices. More specifically, the thermal Kapitza conductance is typically larger for in-plane thermal transport in the presence of scattering interfaces, while a five-fold increase from cross-plane to in-plane directions was observed during coherent heat conduction [53]. To explain this difference, we hypothesize that heat conduction in periodically-twinned Si is subject to a crossover from phonon interface scattering to superlattice-like thermal transport, as the twin period thickness decreases, as discussed below.

In bulk systems, the role of TBs on thermal conductivity can only be associated with anharmonic phonon-phonon interactions [11] or interface scattering effects [20,54]. However, Fig. 3(a) shows that the phonon density of states (PDOS) averaged over all atoms outside TB regions in a twinning superlattice with $2\lambda$ = 3.8 nm, is identical to that of a twin-free $3C$ crystal structure, Fig. 3(b). This result proves that TB effects on anharmonic scattering are negligible even for small twin periods. On the contrary, Fig. 3(a) shows that the twin region defined by atoms within the interface transport length $L_K$ in Fig. 1(c) exhibit different vibrational optical peaks than the pure $3C$ PDOS. In particular, we observe a depression of the high-frequency TO peak, but the low-frequency TA peak remains unchanged. This agrees with the conclusion of Aubry *et al.* [23] showing that TBs easily transmit low-frequency phonons and scatter high-frequency ones.

The inset of Fig. 3(a) shows that the twin region in the above calculation consists of two diamond-hexagonal ($2H$) layers and two $3C$ ones, equivalent to 50% hexagonality [30]. Therefore, to better understand the intrinsic TB contribution on phonons, we examined the PDOS of $2H$ crystals in Si, using MD simulations and *ab-initio* calculations, Fig. 3(b) and Fig. 4, respectively. Interestingly, both semi-empirical and first-principles calculations predict that the PDOS of $2H$ Si polytype and diamond-cubic Si primarily differs on both TA and TO transverse vibrational modes. Phonon dispersion curves obtained by *ab-initio* (Fig. S3) reveal a multiplication of dispersion bands at highest and lowest frequencies in the $2H$ crystals, a phenomenon that is known to introduce significant phonon transport anisotropy compared to the diamond-cubic crystal [55]. A notable property of the 2*H* PDOS is the appearance of two TO peaks, as opposed to one in the 3*C* PDOS. The MD-based PDOS for 4*H* and 6*H* polytypes also present two TO peaks at high frequency, but the behavior at low frequency (TA) is closer to that for the 3*C* PDOS. Therefore, comparing the TA and TO PDOS of atoms in the twin region, Fig. 3(a), to PDOS for different hexagonal Si polytypes, Fig. 3(b), allows us to conclude that the former is closer to the 6*H*-Si PDOS. When the twin period thickness is equal to 1.88 nm (or equivalently TB density = 1.06 nm$^{-1}$), the twinning superlattice structure is matched by the 6*H*-Si polytype. Consequently, the heat transfer mode changes to fully homogeneous, as illustrated in Fig. 3(c).

Moreover, the boundary conductance results in Fig. 1(c), combined with the above PDOS analysis, suggest that the effective cross-plane thermal transport of Si twinning superlattices with a twin period thickness < 22.6 nm is best described in terms of a 3*C*/6*H* heterostructure with no interface scattering, Fig. 3(d). Several assumptions are made in the proposed model: (1) There is no phonon scattering at the interfaces; therefore this assumption only holds for superlattices with 2$\lambda$ < 22.6 nm, (2) the thickness of 6*H* layers representing the TBs is constant and equal to the effective transport length $L_K$ (0.94 nm), and (3) heat conduction is anisotropic and follows Fourier's law in each layer. Thus it is possible to re-write the effective cross-plane thermal conductivity as follows:

$$\kappa_{eff}^{\perp}(\lambda) = \left[ \left(\frac{L_k}{\lambda}\right)\left(\kappa_{6H}^{\perp}\right)^{-1} + \left(1 - \frac{L_k}{\lambda}\right)\left(\kappa_{\lambda_T}^{\perp}\right)^{-1} \right]^{-1}, \qquad (2)$$

where $\kappa_{\lambda_T}^\perp$ represents the minimum of thermal conductivity in the interface-scattering regime, or

$$\kappa_{\lambda_T}^\perp = \left[\frac{1}{\kappa_{3C}^\perp} + \left(\frac{1}{\lambda_T}\right)(\sigma_K)^{-1}\right]^{-1}, \quad (3)$$

with $\lambda_T$ the twin thickness at the interface-scattering to superlattice-like regime transition (= 11.3 nm in this study). Using the computed values for $\kappa_{6H}^\perp$ and $\kappa_{\lambda_T}^\perp$ shown in Table 1, effective thermal conductivities obtained from Eq. (2) and (3) agree well with MD predictions at both 300 K and 700 K for twin period thicknesses smaller than $2\lambda_T$. For the in-plane direction, the 6$H$ and 3$C$ layers represent thermal resistances in parallel; therefore the effective in-plane thermal conductivity of the twinning superlattice is given by:

$$\kappa_{eff}^\parallel(\lambda) = \left(\frac{L_k}{\lambda}\right)\left(\kappa_{6H}^\parallel\right) + \left(1 - \frac{L_k}{\lambda}\right)\left(\kappa_{3C}^\parallel\right) \quad (4)$$

Thermal conductivities predicted from Eq. (4) are plotted in Fig. 2(b) using values in Table I, and also show good match against our MD simulation results.

## C. Thermal conductivity in twinning superlattice nanowires

As we proceed forward from bulk twinning superlattices, we now consider the more common case of semiconductor twinning superlattice NWs [16]. Surface phonon scattering dominates the thermal conductivity of NWs, which was evidenced here by the disappearance of temperature jumps in twinning superlattice NWs containing only one TB (Fig. S4). Studying straight NWs has the advantage that phonon waves are polarized along the main axis [8], so in-plane and cross-plane transport behaviors are more clearly defined.

To study the cross-plane transport, we modeled $[1\bar{1}1]$-oriented NWs with a hexagonal cross-section and smooth sidewalls made of six {110} facets, Fig. 5(a). It is worth noting that phonon backward scattering is greatly reduced with this type of surface facets [7,56,57]. It is shown in Fig. 5(b) that the cross-plane thermal conductivities predicted in NW models exhibit marked reductions as the TB density increases, similar to our bulk models. Apparently, size effects become non-linear for TB densities $\geq 0.53$ nm$^{-1}$, resulting in distinct behaviors depending on the NW length: Thermal conductivities are found to decrease, remain constant or increase for length $L$ = 45 nm, 90 nm, and 150 nm, respectively. In coherent GaAs/AlAs superlattices [9], it was observed experimentally that the thermal conductivity increases with the length of the superlattice structure, which is similar to the present finding for small twin period thicknesses. Therefore, Fig. 5(b) highlights three bulk-like regimes of thermal transport in Si twinning superlattice NWs, from boundary scattering to superlattice-like and fully homogeneous regimes.

Furthermore, the in-plane direction was studied using NW models with lengthwise twins oriented along a <112> direction with a square cross-section made of $(1\bar{1}1)$ and {110} sidewalls [17,21]. The models with square cross-section were constructed in such a way that the cross-sectional area was as close as possible to that of the NWs with hexagonal cross-section. Figs. 6(a), (b) show that thermal conductivity of lengthwise twinned NWs is almost constant and equal to the $6H$ predictions for the same NW geometry, 11.3 W/mK. This confirms that heat conduction in lengthwise NWs is also similar to in-plane transfer through a $3C/6H$ heterostructure. Furthermore, Fig. 6(c) shows the PDOS averaged over all atoms present in the core or on the {110} surface sidewalls of a <112> oriented square NW. Note that only analysis on {110} sidewalls is relevant in this case because {111} sidewalls are not intersected by defects. In the twin-free NW, PDOS curves between surface and core atoms mainly differ by a shift to lower frequencies of the TA peak, which has commonly been attributed to phonon confinement in Si NWs [58], and a large depression of the TO peak. The addition of nanotwins ($2\lambda = 3.8$ nm) in Fig. 6(c) dramatically reduces the height of the high-frequency TO peaks, but with the same magnitude for both core and surface atoms. Therefore this result proves that nanotwin effects in straight NWs are independent of surface phonon scattering processes, but only relates to bulk scattering; i.e., the intrinsic influence of nanotwins on thermal conductivity is governed by the same underlying physical process in bulk and NW Si.

## IV. CONCLUSION

MD simulations and *ab-initio* calculations have shown evidence for intrinsic reductions of cross-plane and in-plane thermal conductivities in bulk Si and Si NWs, as the twin period thickness decreases at the nanoscale. This phenomenon arises from a regime transition from boundary-scattering to superlattice-like heat conduction. Modeling TBs as atomically-thin 6*H*-Si layers, rather than phonon scattering interfaces, provides an accurate description of the effective thermal conductivity in twinning superlattices with twin period thickness smaller than 22.6 nm. Furthermore, superlattice-like phonon transport was found over a wide range of twin period thicknesses, for both in-plane and cross-plane directions, which holds promise for minimizing the thermal conductivity when the twin period thickness is not uniform. These findings have significant implications for extending the use of superlattices in nanoscale thermoelectrics and optoelectronics by designing materials with coherent interfaces.

**ACKNOWLEDGMENTS**

The authors acknowledge the resources of the Extreme Science and Engineering Discovery Environment (XSEDE) supported by National Science Foundation grant number OCI-1053575, and the Vermont Advanced Computing Core supported by NASA (NNX-08AO96G), as well as support from NASA grant NNX14AN20A and NSF-REU grant 1062966.

**SUPPLEMENTAL MATERIALS AVAILABLE:**

Atomistic modeling of 3*C*, 6*H*, 4*H* and 2*H* Si crystals. Supporting Figures S1 – S4.

**Table I. Cross-plane and in-plane thermal conductivities for bulk 3*C* and 6*H* crystals (*L* = 90 nm).**

| Temperature (K) | Transition $\lambda_T$ (nm) | $\kappa^\perp_{\lambda_T}$ (W/mK) | $\kappa^\perp_{3C}$ (W/mK) | $\kappa^\perp_{6H}$ (W/mK) | $\kappa^\parallel_{3C}$ (W/mK) | $\kappa^\parallel_{6H}$ (W/mK) |
|---|---|---|---|---|---|---|
| 300 | 11.3 | 14.6 | 20.6 | 16.5 | 19.8 | 15.6 |
| 700 | 11.3 | 12.5 | 15.4 | 11.6 | - | - |

**FIGURE CAPTIONS**

**FIG 1. (Color online) Influence of twin period thickness on Kapitza boundary conductance in bulk Si containing periodic nanotwins by non-equilibrium MD simulation.** (a) Atomistic model for a $[1\bar{1}\bar{1}]$-oriented bulk Si crystal of twin period thickness $2\lambda$. (b) Temperature profile across a single twin boundary superimposed on the atomic structure at the interface shown in dark red color. A change of temperature $T_k$ across the interface is clearly visible. Blue lines represent best fit lines used to calculate the thermal conductivity of the grain region. (c) Twin boundary thermal conductance as a function of twin period thickness $2\lambda$ for two system lengths and two heat fluxes.

**FIG 2. (Color online) Simulated bulk thermal conductivities as a function of twin boundary density.** (a) Cross-plane thermal conductivity through coherent twinning superlattices. Dashed lines represent the diffusive boundary scattering regime from Eq. (1). Solid lines represent predictions from PDOS-based model proposed for semi-coherent phonon transport according to Eq. (2) – (3). (b) In-plane thermal conductivity parallel to lengthwise twins. The solid line represents predictions from Eq. (4). All systems have same length $L = 90$ nm.

**FIG 3. (Color online) Phonon density of states (PDOS) in cross-plane heat conduction of bulk Si systems by MD simulation at 300 K.** (a) PDOS for atoms in non-twin and twin regions in a Si twinning superlattice with twin period thickness of 3.8 nm. The structure of each slide used in the PDOS calculations is shown in inset. (b) PDOS of different single-crystalline Si polytypes (3$C$, 6$H$, 4$H$ and 2$H$) with heat flux imposed along the [111] direction used for comparison. The frequency domains for acoustic (TA, LA) and optical (TO, LO) modes are indicated. Proposed models for the thermal transport behavior of (c) homogeneous, (d) superlattice-like and (e) boundary-scattering twinning superlattices.

**FIG 4. (Color online)** DFT-based ab-initio calculation of phonon density states in 3$C$ Si and 2$H$ Si.

**FIG 5. (Color online) Cross-plane thermal conductivity of twinning superlattices in single Si NWs.** (a) NW geometry oriented along [111] direction with straight sidewall made of six {110} planes. (b) Non-equilibrium MD computations with different NW lengths, as a function of TB density.

**FIG 6. (Color online) In-plane thermal conductivity of twinning superlattices in single Si NWs.** (a) NW geometry oriented along [112] direction with straight sidewalls made of {110} and {111} planes. (b) Non-equilibrium MD computations with $L = 90$ nm, as a function of TB density. Associated PDOS spectrum for core and surface atoms in (c) twin-free and (d) nanotwinned Si NWs.

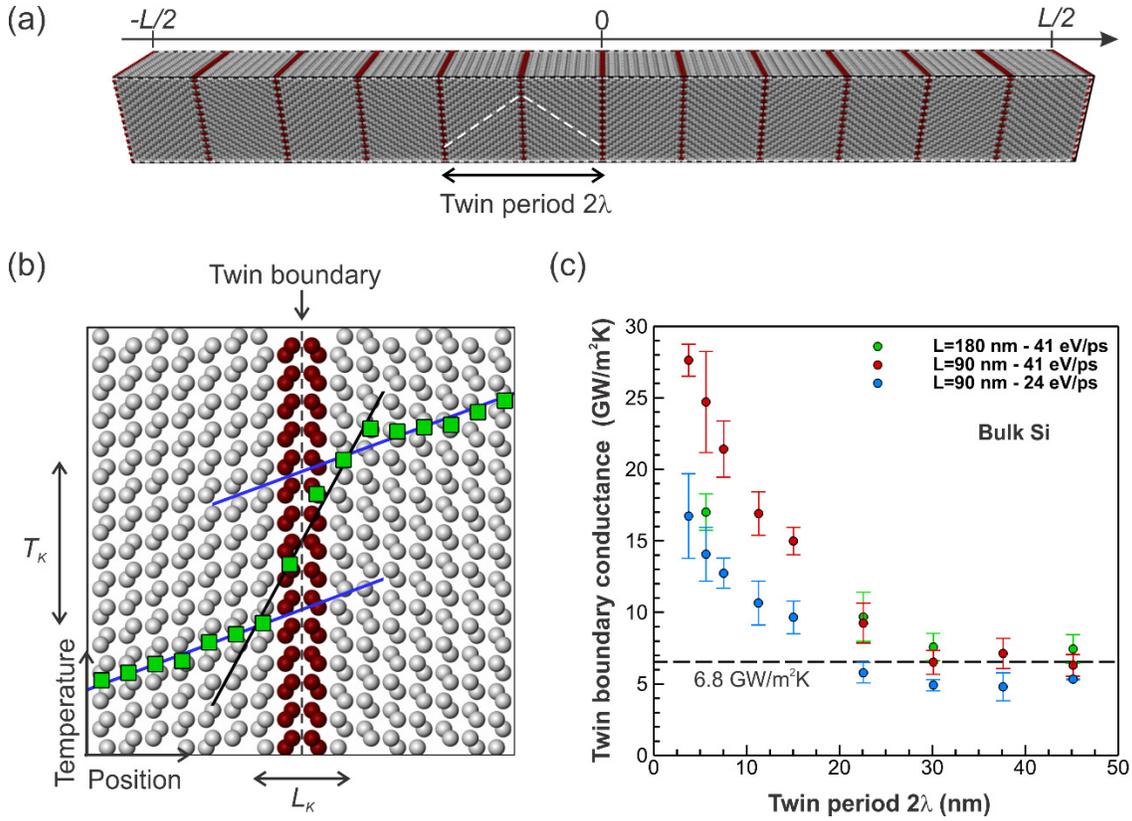

**FIG 1. (Color online) Influence of twin period thickness on Kapitza boundary conductance in bulk Si containing periodic nanotwins by non-equilibrium MD simulation.** (a) Atomistic model for a $[1\bar{1}1]$-oriented bulk Si crystal of twin period thickness $2\lambda$. (b) Temperature profile across a single twin boundary superimposed on the atomic structure at the interface shown in dark red color. A change of temperature $T_k$ across the interface is clearly visible. Blue lines represent best fit lines used to calculate the thermal conductivity of the grain region. (c) Twin boundary thermal conductance as a function of twin period thickness $2\lambda$ for two system lengths and two heat fluxes.

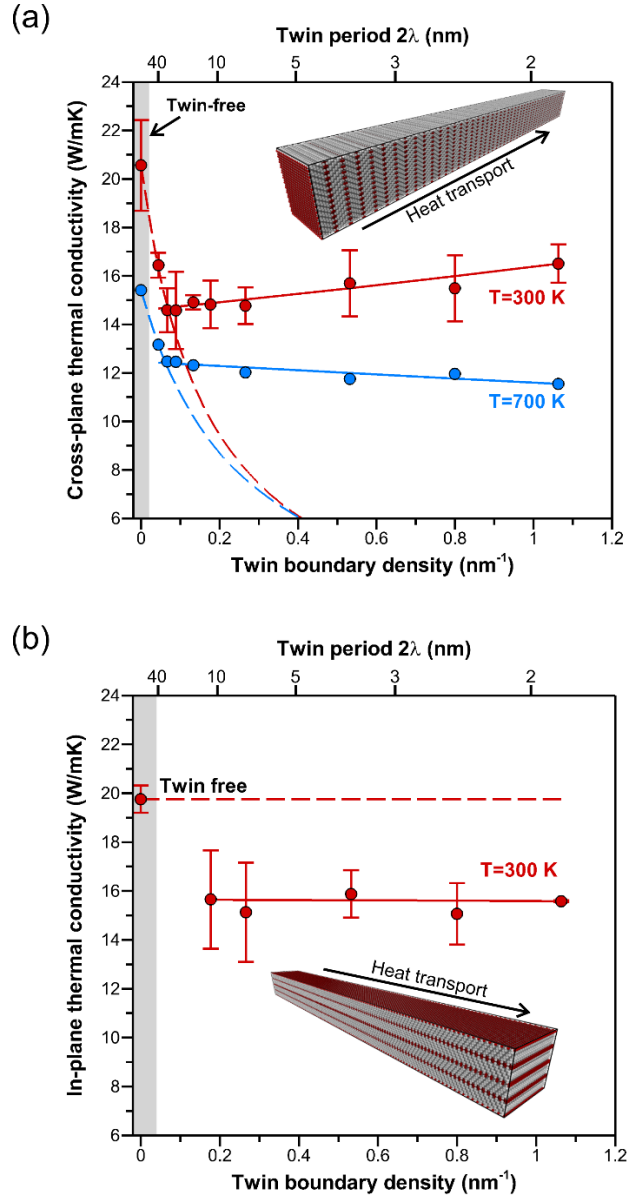

**FIG 2. (Color online) Simulated bulk thermal conductivities as a function of twin boundary density.** (a) Cross-plane thermal conductivity through coherent twinning superlattices. Dashed lines represent the diffusive boundary scattering regime from Eq. (1). Solid lines represent predictions from PDOS-based model proposed for semi-coherent phonon transport according to Eq. (2) – (3). (b) In-plane thermal conductivity parallel to lengthwise twins. The solid line represents predictions from Eq. (4). All systems have same length $L = 90$ nm.

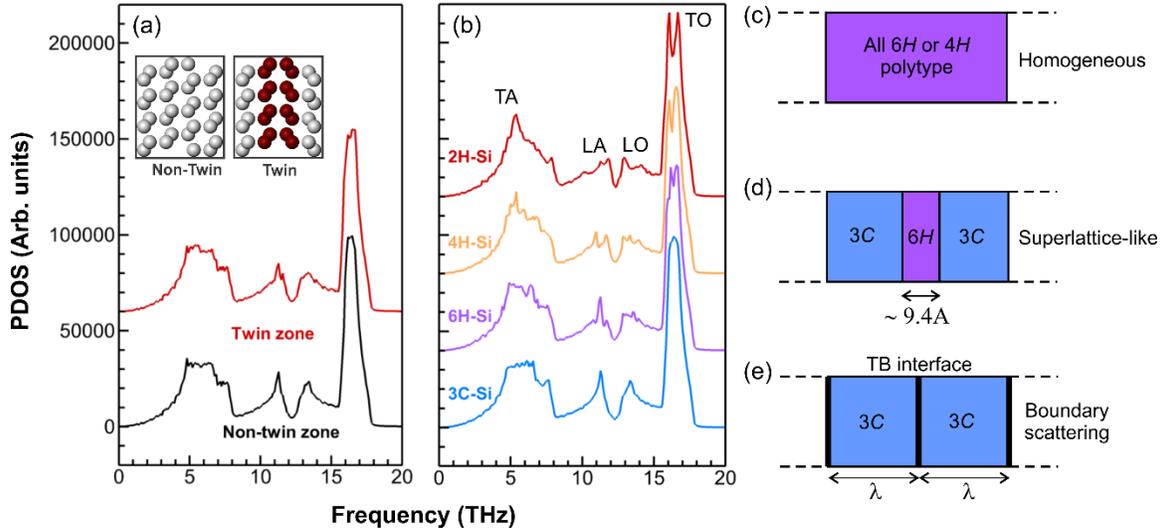

**FIG 3. (Color online) Phonon density of states (PDOS) in cross-plane heat conduction of bulk Si systems by MD simulation at 300 K.** (a) PDOS for atoms in non-twin and twin regions in a Si twinning superlattice with twin period thickness of 3.8 nm. The structure of each slide used in the PDOS calculations is shown in inset. (b) PDOS of different single-crystalline Si polytypes (3*C*, 6*H*, 4*H* and 2*H*) with heat flux imposed along the [111] direction used for comparison. The frequency domains for acoustic (TA, LA) and optical (TO, LO) modes are indicated. Proposed models for the thermal transport behavior of (c) homogeneous, (d) superlattice-like and (e) boundary-scattering twinning superlattices.

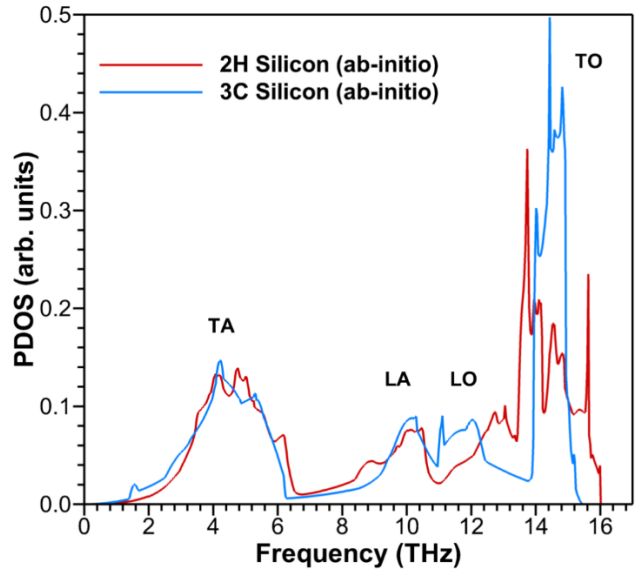

**FIG 4. (Color online)** DFT-based ab-initio calculation of phonon density states in 3*C* Si and 2*H* Si.

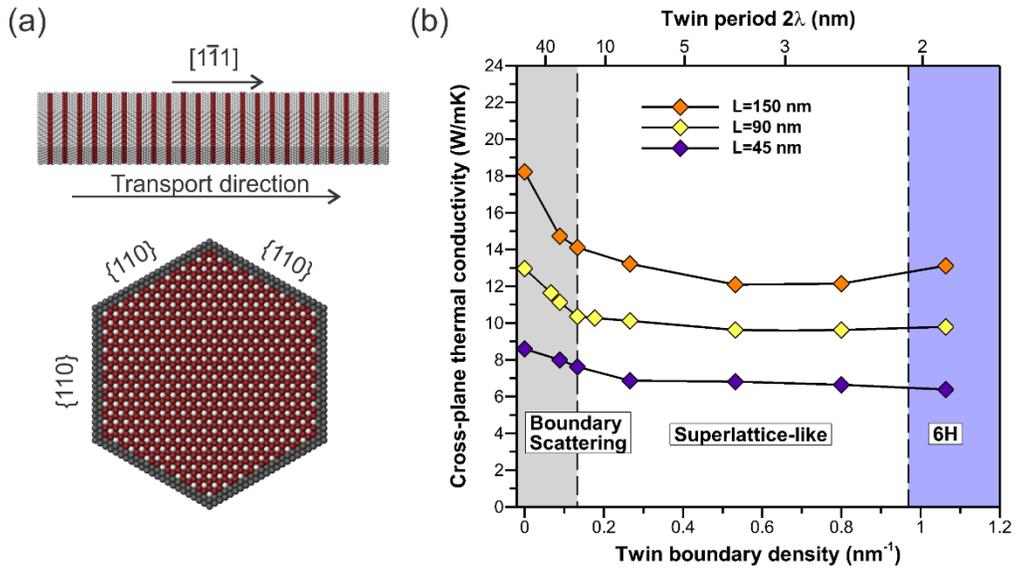

**FIG 5. (Color online) Cross-plane thermal conductivity of twinning superlattices in single Si NWs.** (a) NW geometry oriented along [111] direction with straight sidewall made of six {110} planes. (b) Non-equilibrium MD computations with different NW lengths, as a function of TB density.

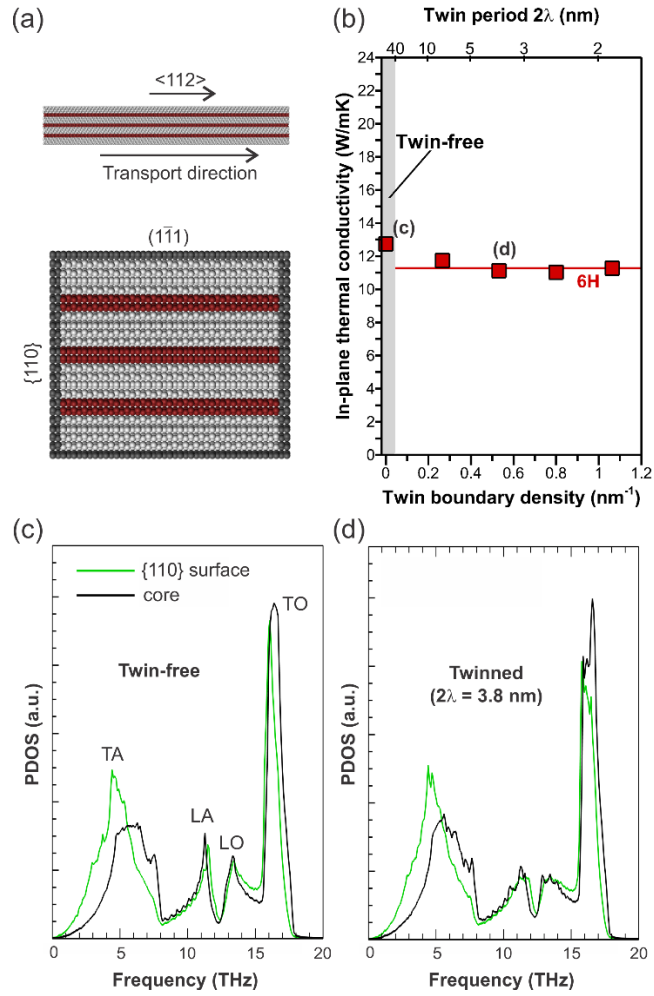

**FIG 6. (Color online) In-plane thermal conductivity of twinning superlattices in single Si NWs.** (a) NW geometry oriented along [112] direction with straight sidewalls made of {110} and {111} planes. (b) Non-equilibrium MD computations with $L = 90$ nm, as a function of TB density. Associated PDOS spectrum for core and surface atoms in (c) twin-free and (d) nanotwinned Si NWs.